\documentstyle{article}

\newcommand{\mathbf}{\bf}

\begin{document}

\begin{center}
{\huge\bf On Quantum Mechanics}
\end{center}

\vspace{1cm}
\begin{center}
{\large\bf
F.GHABOUSSI}\\
\end{center}

\begin{center}
\begin{minipage}{8cm}
Department of Physics, University of Konstanz\\
P.O. Box 5560, D 78434 Konstanz, Germany\\
E-mail: ghabousi@kaluza.physik.uni-konstanz.de
\end{minipage}
\end{center}

\vspace{1cm}

\begin{center}
{\large{\bf Abstract}}
\end{center}

\begin{center}
\begin{minipage}{12cm}
We discuss the axiomatic basis of quantum mechanics and show that  
it is neither general nor consistent, since its axioms are  
incompatible with each other and moreover it does not incorporate  
the magnetic quantization as in the cyclotron motion. A general and  
consistent system of axioms is conjectured which incorporates also  
the magnetic quantization.

\end{minipage}
\end{center}

\newpage
The neccessity of quantum mechanics (QM) was a result of  
experimental data for the energy spectrum of atoms which could not  
be explained in accord with the laws of classical mechanics.  
Nevertheless in the axiomatic basis of QM there are axioms which are  
not supported by the experimental results, since the whole system  
of axioms (1) is introduced {\it only} as a "plausible  
generalization" of the first axiom \cite{a}:

\begin{equation}
[\hat{P}_i \ \ , \ \ \hat{Q}_j ] = - i \hbar \delta_{ij} \ \ \ , \  
\ \  [\hat{P}_i \ \ , \ \ \hat{P}_j ] = 0 \ \ \ , \ \ \ [\hat{Q}_i \  
\ , \ \ \hat{Q}_j ] = 0 \ \ \,
\end{equation}

where $i, j = 1, ..., k$ and $\hat{P}_i$ and $\hat{Q}_i$ are the  
momentum and position operators of a quantum system with $k$ degrees  
of freedom. Note as an importent fact about the application of QM  
that the usual application of the QM to the energy spectrum of atoms  
needs only to use the first axiom to quantize the Hamilton  
operator, but it does not need any use of the other two axioms  
\cite{power}. Thus these two axioms are not involved yet in any  
quantum theory. Hence, in view of the fact that the energy  
quantization was the only application of the system (1), therefore  
the two last axioms remained without application and it was not  
possible to prove their compatibility with the first axiom within a  
concrete physical question.

Nevertheless since these axioms, as the first principles of QM, are  
based as usual on plausible arguments, but not on other quantum  
axioms or empirical results. Therefore one can not exclude  
inconsistencies within the system (1) a periori, so that a general  
revision of the axiomatic structure of QM seems to be neccessary.  
Thus the appearance of new quantum effects like the magnetic  
quantization in the cyclotron motion enforces such a revision,  
specially if one will describe them in accord with QM. In other  
words, among others, a reason to revise the axiomatic basis of QM is  
that the quantum commutator in the two dimensional cyclotron motion  
\cite{aoki}:
\begin{equation}
e B [\hat{Q}_m \ \ , \ \ \hat{Q}_m ] = - i \hbar \epsilon_{mn} \ \  
\  ; m, n = 1, 2 \ \ \ , \ \epsilon_{mn} = - \epsilon_{nm} = -1 \ \  
\ ,
\end{equation}

with $Q_m$ as the relative coordinates of an electron moving in a  
constant magnetic field $B$, is not compatible with the system (1).

From dimensional analysis it is obvious that since the action  
$(\sim \hbar)$ has the dimension of momentum times position, i. e.:  
$P \cdot Q = L^{-1} \cdot L = L^0$ in geometric units.
Hence the dimensionality is saved in this case, in view of $dim B  
\sim L^{-2}$. Therefore also this commutator should define, in  
principle, a quantum postulate. But if one considers the commutator  
(2) as a quantum postulate, then the system (1) can be considered  
neither as the general basis for quantum postulates nor as a  
consistent algebra.

A more conceptual reason to neglect the cyclotron commutator, as a  
possible quantum postulate, was the common but a periori belive that  

"particles" and "fields" should have two fundamentaly different  
nature and so they should obey two different type of mechanics or  
dynamics. Therefore the quantum commutator postulates of particles  
and fields should be different and without any common relation. Even  
after the rise of quantum field theory, where quantized fields  
appear as particles and particles appear as quantum fields, the main  
difference still remained. Although for example the quantum  
electrodynamics (QED) is based on the equivalence of the quantum  
behaviour of electrons and electromagnetic potential field  
\cite{heit}.

In accord with QED, in the same manner that quantum mechanical  
properties of the charged test body ( $\sim$ electron ) prevent an  
exact measurment of the electromagnetic field, the quantum  
electrodynamical properties of the electromagnetic field prevent an  
exact measurment of the position of the charged test body  
\cite{heit}. In other words the uncertainties of electron causes the  
uncertainties of the electromagnetic field and vice versa.
Thus, in accord with QM where the quantum character of a particle  
is manifested for example by its uncertainty relations, the quantum  
character of electron depends on the existence of the uncertainty  
relations of the electromagnetic field, which manifest the measuring  
interaction between the electron and the field \cite{heit}.
Hence QED (of electromagnetic fields ) and the QM ( of electrons )  
are two inseparable part of a quantum theory ( of particles and  
fields ) and neither is consistent without the other \cite{heit}.

Nevertheless, it was the mentioned artificial difference between  
fields and particles with its further consequences which prevented  
to interpret the phenomenologically introduced commutator in the  
cyclotron motion, i. e.: $e B [\hat{Q}_i \ \ , \ \ \hat{Q}_j ] = - i  
\hbar$, as what it is, namely as a quantum commutator postulate:  
Since in this case the position operators of electron and the  
magnetic field $B$ appear together in one and the same relation.

The internal incompatibility of the algebra (1) is based, besides  
of this fact, on the disconnectedness of its axioms, since they are  
assumed just as a "plausible generalization" [1] partly as the  
quantum property and partly as the classical properties of a free  
system: Thus by the last two axioms, the quantum behaviour of, e. g.  
a quantum particle, is considered to be the same as the classical  
properties of a classical particle in a free motion. Thus even the   
obvious relations between the momentum and the position coordinates  
of a classical particle in a bounded motion are ignored, where for  
example $P_m = M \dot{Q}_m = \epsilon_{mn} M Q_n \dot{\alpha}$ for  
$Q_1 = r cos \alpha, Q_2 = r  sin \alpha$. In {\it this case} the  
Poisson brackets:

${\{ P_1 \ \ , \ \ P_2 }\} \propto {\{ P_1 \ \ , \ \ Q_1 }\}$ and  
${\{ Q_1 \ \ , \ \ Q_2 }\} \propto {\{ Q_1 \ \ , \ \ P_1 }\}$ are  
non-trivial \cite{nn}. In other words, in the case of bounded  
motion, the non-triviality of the first bracket of Poisson algebra  
requires also the non-trivialities of the second and the third  
brackets. Thus in view of the mentioned correspondance between  
Poisson- and the Heisenberg algebra, the same requirement should  
hold also for the corresponding commutators in the system (1) for  
the related quantum case.

To prove the general contradiction between the axioms of the system  
(1) let us consider the most minimal case where the algebra (1) can  
be proved, namely for $i, j = m, n = 1, 2$. We will prove that the  
first axiom of the system (1) is incompatible with the rest two  
axioms: This is obvious implicitely from the well known relation  
between the phase space variables and their quantum operators, in  
accord to which if two such operators commute with each other, then  
one of these variables is a function of the other one \cite{fun}. On  
the other hand if two such variables are independent of each other,  
then the commutator of their operators need not to vanish, which is  
obvious from the first axiom of the system (1). These relations  
follows from the mentioned correspondance btween the commutators and  
Poisson brackets. Thus the asumption of the first commutator in  
(1), in accord to which $\hat{P}_m$ commute with $\epsilon_{mn}  
\hat{Q}_n$, but not with $\hat{Q}_m$, means that $P_m$ is a function  
of $\epsilon_{mn} Q_n$, but not of $Q_m$, i. e.: $P_m = f  
(\epsilon_{mn} Q_n)$ and also $P_m \neq f (Q_m)$. Hence in view of  
$Q_m \neq f (\epsilon_{mn} Q_n)$, it follows that $P_m \neq f (P_n)$  
and therefore the related operators need not to commute, i. e.: $[  
\hat{P}_m \ , \ \hat{P}_n ] \neq 0$ \cite{con}. Note that these  
arguments about the relation between $P_m, Q_m$ and their operators  
is not in contradiction with the Hamitonian case where the  
Hamiltonian $H$ is a quadratic function of $P_m$ and $Q_m$  
variables, but its operator $\hat{H}$ does not commute with those of  
$\hat{P}_m$ and $\hat{Q}_m$. Since, not only that in the above  
discussed case $P_m$ are only linear functions of $\epsilon_{mn}  
Q_n$, whereas the Hamiltonian is quadratic function of them, but  
also the direction of conclusion does not contradict the Hamiltonian  
case. A contradiction with the Hamiltonian case would appear, if we  
required the oposite direction of conclusion (see \cite{con})  
\cite{cc}.

To prove this fact explicitely let us consider the wave function of  
the system (1), with respect to which the commutators of (1) can be  
proved directly, to be in the position representation, i. e.:
$\Psi ((1)) := \Psi (Q_1, Q_2)$. Hence the momentum operators  
should act as differential operators, i. e.:
$\hat{P}_m \Psi (Q_1, Q_2) = -i \hbar \partial_m \Psi (Q_1, Q_2) =  
P_m \Psi (Q_1, Q_2)$ and the position operators should act by  
multiplication, i. e.:

$\hat{Q}_m \Psi (Q_1, Q_2) = Q_m \Psi (Q_1, Q_2)$. On the one hand  
the assumption of the first postulate in (1), i. e.: $[\hat{P}_1 \ \  
, \ \ \hat{Q}_1 ] \neq 0$, $[\hat{P}_2 \ \ , \ \ \hat{Q}_2 ] \neq  
0$, $[\hat{P}_1 \ \ , \ \ \hat{Q}_2 ] = 0$ and $[\hat{P}_2 \ \ , \ \  
\hat{Q}_1 ] = 0$ demands that, as we discussed above, the momentum  
variables $P_m$ can not be functions of position variables $Q_m$ but  
they must be functions of $\epsilon_{mn} Q_n$ variables, i. e.:  
$P_m = f (Q_n \epsilon_{mn})$. A dependence between momentum- and  
position variables which is similar to the above introduced example  
of bounded motion. On the other hand, if so then the second  
commutator of the standard quantum algebra (1) is not more fulfilled  
in this case, since as in the bounded motion, this commutator is  
not trivial for the case of a linear dependence: $P_m \propto  
\epsilon_{mn} Q_n$. Thus $[\hat{P}_1 \ \ , \ \ \hat{P}_2 ] \Psi  
(Q_1, Q_2) \neq 0$ by calculation. Hence the contradiction between  
the first and the second commutators in the algebra (1) can be  
proved also explicitely: Thus in the quantized bounded motion which  
is similar to the cyclotron motion of electron in a magnetic field,  
the second commutator is given by: $[\hat{P}_1 \ \ , \ \ \hat{P}_2 ]  
\Psi (Q_1, Q_2) = \hat{P}_1 ( P_2 \cdot \Psi ) - \hat{P}_2 ( P_1  
\cdot \Psi ) = - 2 i \hbar M \dot{\alpha} \Psi (Q_1, Q_2)$ in  
contrast to the system (1), since $P_m = \epsilon_{mn} M \cdot Q_n  
\cdot \dot{\alpha}$ and $\hat{P}_n = - i \hbar \partial_n$.  
Therefore in view of the fact that in the case of electromagnetic  
interaction the dimensionless electron charge $e$ should be involved  
as a coupling constant, one may set for the cyclotron motion:  
$\dot{\alpha} = \omega_c = \displaystyle{\frac{e B}{2 M_e}}$. So  
that for the electron coupled to the electromagnetic field $B$, the  
commutator: $[\hat{P}_1 \ \ , \ \ \hat{P}_2 ]$ results in:

\begin{equation}
[\hat{P}_1 \ \ , \ \ \hat{P}_2 ] = i \hbar e B \ \ ,
\end{equation}

which is in contradiction of with the second axiom in (1).

One can even prove that the third commutator in the system (1)  
results, for the same quantized bounded motion where the position  
operators in the momentum representation are given by: $\hat{Q}_m =   
i \hbar \partial_{P_m}$, in: $[\hat{Q}_m \ \
 , \ \ \hat{Q}_n ] \Psi (P_1, P_2) = - 2 i \hbar (M  
\dot{\alpha})^{-1} \Psi (P_1, P_2)$. Thus it can be rewritten by the  
commutator (2) for the case of cyclotron motion, again in accord  
with $\dot{\alpha} = \omega_c = \displaystyle{\frac{e B}{2 M_e}}$.

Another conceptual basis to chooose the algebra (1) was also the a  
periori concept of "free" quantum particle, e. g. an "electron"  
without interaction with any field, thus such a free particle have  
commuting position- and also momentum operators. Nevertheless as it  
is known from QED \cite{heit}, such a "free" electron does not  
exists within the context of QED, since as it is discussed above an  
"electron" without interaction with quantized electromagnetic field  
can not be considered as a quantum particle: Thus, in accord with  
Heisenberg's argument, in order that the uncertainty relations  
$\Delta P_i \cdot \Delta Q_i \geq \hbar$ are given for an electron  
as a quantum particle, there must be given an uncertainty relations  
for the measurment of electron by an electromagnetic field in accord  
with: $\Delta G_i \cdot \Delta Q_i \geq \hbar$, where $G_i$ is the  
field momentum of the observing
electromagnetic field \cite{heit}. In other words the measurment or  
interaction of an electron with the electromagnetic field, which is  
manifested by the last uncertainty relation, is the presupposition  
for quantum character of electron which is manifested by the  
uncertainty relation $\Delta P_i \cdot \Delta Q_i \geq \hbar$. Hence  
the existence of uncertainty relation $\Delta P_i \cdot \Delta Q_i  
\geq \hbar$ depends on the existence of the uncertainty relations  
$\Delta G_i \cdot \Delta Q_i \geq \hbar$ which manifests the  
interaction between the electron and the electromagnetic field.  
Therefore in view of the QM fact that the existence of uncertainty  
relations is equivalent to the existence of related commutators, the  
discussed interaction between electron and the electromagnetic  
field is the presupposition for the correctness of the first  
commutator in (1). Hence a system of axioms which contains the first  
commutator in (1), can not apply to a "free" electron, but it  
applies to a particle with electromagnetic interaction. As a first  
consequence, the system (1) which presuppsoes the existence of free  
quantum particle is inconsistent, as we showed above implicitely and  
explicitely. Moreover in the absence of such a "free" motion, it is  
no neccessity to assume the second and the third commutators in (1)  
which manifest the free motion \cite{conj}, but one should  
postulate other axioms which are suitable for a bounded motion.  
Nevertheless we will show that the electron as a quantum particle,  
not only in the cyclotron motion, but in view of its general  
neccessary interaction with the electromagnetic field which  
manifests the quantum character of electron, does not obey the  
system of axioms (1), but it should obey a system of axioms with  
non-trivial second and third commutators. This system will be the  
general and consistent one for a quantum particle like electron,  
since despite of the system (1) it considers the neccessary coupling  
of electron, as a quantum particle, to the electromagnetic field.  
Therefore it will be also the quantum algebra of quantum  
electrodynamical effects of electron, like the cyclotron motion and  
the flux quantization.

To prove this, first note that the field momentum of  
electromagnetic field: $G_i = \int \epsilon_{i j k} E_j B_k d^3 x$;

$i, j, k = 1, 2, 3$ is equal to $e A_i$ where $A_i$ is the  
electromagnetic potential. This equality can be derived for $E_j$  
and $B_k$ as the solutions of the inhomogeneous Maxwell equations  
for an electromagnetic field coupled to a single electron. If one  
uses the Gauss' law for $E_j$ and the integral $A_i = \int  
\epsilon_{i j k} B_k dx_j = \epsilon_{i j k} B_k x_j$, in view of  
$div B = 0$. Hence the above introduced uncertaity relation $\Delta  
G_i \cdot \Delta Q_i \geq \hbar$ can be rewritten by $e \Delta A_i  
\cdot \Delta Q_i \geq \hbar$ which should be considered as the  
presupposition for the quantum character of electron.

\bigskip
The argument to prove the general neccessity of non-trivial  
commutators for a quantum particle like elektron, is based on the  
fact that on the one hand the quantized electromagnetic potential,  
the photon, possess two degrees of freedom or two components $A_m$  
which are given in two dimensions by

$A_m = B \cdot Q^n \epsilon_{mn}$ \cite{grads}. On the other hand  
in accord with the above analysis the quantum character of electron  
which is manifested by its uncertainty relations, presupposes the  
uncertainty relations

$e \Delta A_i \cdot \Delta Q_i \geq \hbar$. Therefore in view of  
the fact that such an interaction, to determine the position of  
electron in the $Q_1$- direction, causes also an uncertainty $\Delta  
Q_2$ in the position
of electron in the $Q_2$- direction, in accord with: $\Delta A_1 =  
B \cdot \Delta Q_2$ and $e \Delta A_1 \Delta Q_1 = e B \Delta Q_2  
\Delta Q_1 \geq \hbar$. Hence in view of the QM equivalence between  
commutators and related uncertainty relations, the existence of the  
uncertainty relation $ e B \Delta Q_1 \Delta Q_2 \geq \hbar$ for an  
electromagnetically measured electron is equivalent to the existence  
of the commutor (2), i. e. $e B [\hat{Q}_i \ \ , \ \ \hat{Q}_j ] =  
- i \hbar$, for the electron as a quantum particle. Then this  
commutator should replace the third commutator in (1). Thus in  
accord with this replacement and the above analysis of the measuring  
interaction between electron and electromagnetic field which  
results in commutator (3), also the second commutator in (1) should  
be replaced by the commutator (3).

In other words the new general and consistent system of axioms are  
given by: $(i,j = m, n)$

\begin {equation}
[\hat{P}_m \ \ , \ \ \hat{Q}_n ] = - i \hbar \delta_{mn} \ \ \ , \  
\ \  [\hat{P}_m \ \ , \ \ \hat{P}_n ] = i \epsilon_{mn} \hbar e B \  
\ \ , \ \ \ [\hat{Q}_m \ \ , \ \ \hat{Q}_n ] = - i \epsilon_{mn}  
\hbar (e B)^{-1}
\end{equation}

To see the consistency of this system of axioms, note that  
considering the quantization condition $P_m = e A_m$ which is used  
also in the flux quantization \cite{qq}, these tree commutators are  
equivalent to each other by: $P_m = e A_m = e B \cdot Q^n  
\epsilon_{mn}$. In other words one can consider the algebra (4) as  
various representations of one and the same commutator:

\begin {equation}
 [\hat{Q}_m \ \ , \ \ \hat{Q}_n ] = - i \epsilon_{mn} \hbar (e  
B)^{-1}  \ \ \ , \ \ \ B \cdot \hat{Q}_m = \hat{P}_n \epsilon_{nm}
\end{equation}

In conclusion let us denote that the classical limit of cyclotron  
motion, i. e. the $B \rightarrow 0$ limit is equivalent to the  
classsical limit: $\hbar \rightarrow 0$ where the area and the  
radious of motion surface become very large, i. e. close to the  
rectilinear motion which can be considered as a bounded motion with  
an infinite large radious and area. Moreover note that the algebra  
(5), i. e.: $ e B \epsilon_{nm} [\hat{Q}_m \ \ , \ \ \hat{Q}_n ] = -  
i \hbar$, describes beyond the cyclotron motion also the flux  
quantization which is given usually by $e \int \int F_{mn} dQ^m  
\wedge dQ^n =$

$e \epsilon_{mn} B \int \int dQ^m \wedge dQ^n = e \oint A_m d Q^m =  
N h \ \ , N \in {\mathbf Z}$ for a constant magnetic field $B$.  
Since in view of $e \epsilon_{mn} B \int \int dQ^m \wedge dQ^n = e  
\oint \epsilon_{mn} B \cdot Q^m \wedge dQ^n$ and in accord with the  
QM equivalence between the two quantization postulates in the  
canonical quantization, i. e. $\int \int dP_m \wedge \d Q^m = \oint  
P_m dQ^m = Nh$ and $[\hat{P}_m \ \ , \ \ \hat{Q}_n ] = - i \hbar  
\delta_{mn}$, the integral form of flux quantization relation is  
equivalent to the quantum commutator axiom: $ e B \epsilon_{mn}  
[\hat{Q}_m \ \ , \ \ \hat{Q}_n ] = - i \hbar$. Thus a comparison  
between the mentioned canonical quantization integrals and the flux  
quantization integrals manifests beyond the flux quantization  
requirement \cite{qq} also the neccesity of relation $P_m = e A_m =  
e B \cdot Q^n \epsilon_{mn}$.

\bigskip
\bigskip
Footnotes and references


\begin{thebibliography}{100}
\bibitem{a}
Note that the system (1) is introduced as the "plausible  
generalization" of the first axiom: $[\hat{P}_i \ \ , \ \ \hat{Q}_j  
] = - i \hbar \delta_{ij}$ which is the main axiom of the system  
(1). See: M. Born, W. Heisenberg, P. Jordan, Z. Phys. 35, 557  
(1926).
\bibitem{power}
In other words there are no terms of the: $[\hat{P}_i \ \ , \ \  
\hat{P}_j ]$ or $[\hat{Q}_i \ \ , \ \ \hat{Q}_j ]$ form in the  
energy quantization.
\bibitem{aoki}
H. Aoki: Rep. Prog. Phys. 50, (1987), 655.
Note that also in the Landau model of magnetic quantization only  
the first axiom in (1) is used to quantize the energy, since for  
this one needs only the first axiom. Thus the cyclotron commutator  
is not considered there as a canonical commutator postulate in the  
sense that we do, but it was considered only as a phenomenological  
relation.
\bibitem{nn}
Recall that the quantum Heisenberg algebra (1) corresponds to the  
classical Poisson algebra ${\{ P_i \ \ , \ \ Q_j }\} = \delta_{ij} \  
\ , \ \ {\{ P_i \ \ , \ \ P_j }\} = {\{ Q_i \ \ , \ \ Q_j }\} = 0$.  
Note that the previous dimensional considerations are valid also  
here.
\bibitem{fun}
In other words the conclusion is: $[ \hat{A} \ \ , \ \ \hat{B} ] =  
0 \ \rightarrow A = f (B)$. Note that the oposite conclusion, i. e.:  
$C = f (D) \not\rightarrow [ \hat{C} \ \ , \ \ \hat{D} ] = 0$ is  
not intended here. In view of the fact that the axioms of QM are  
given as commutators of operators, therefore one needs here only the  
first conclusion.
\bibitem{con}
In other words the direction of conclusion is: $P_m \neq f (P_n )  
\rightarrow [ \hat{P}_m \ , \ \hat{P}_n ] \neq 0$. The oposite  
direction will be: $[ \hat{A} \ , \ \hat{B} ] \neq 0 \rightarrow A  
\neq f (B )$, which is not intended here.
\bibitem{conj}
The free motion or the motion of a free particle is the motion of a  
particle that stands under no potential or no force or no field  
strength. In other words for a free motion one has $B = 0$  
everywhere.
\bibitem{qq}
Note that the flux quantization requirement of vanishing of current  
density in the {\it contour} region is equivalent to the vanishing  
of the velocity of electrons in this region which is also equivalent  
to $P_m = e A_m$ in this region, since: $V_{(contour)} =  
\displaystyle{\frac{1}{m_e}} ( P_m - e A_m )_{(contour)} = 0$.
\end{thebibliography}
\end{document}